\documentclass[twocolumn]{aastex631}
\usepackage{hyperref}
\usepackage{color}
\usepackage{amssymb}
\usepackage{amsmath}
\usepackage[ruled,linesnumbered]{algorithm2e}
\usepackage{footmisc}
\usepackage{ulem}

\shorttitle{GesaRaT}
\shortauthors{Chen et al.}
\begin{document}

\title{SEDONA-GesaRaT: an AI-Accelerated Radiative Transfer Program for 3-D Supernova Simulations}
\correspondingauthor{Ulisses Braga-Neto}
\email{ulisses@tamu.edu}

\author[0000-0003-3021-4897]{Xingzhuo Chen}
\affiliation{George P. and Cynthia Woods Mitchell Institute for Fundamental Physics \& Astronomy, \\
Texas A. \& M. University, Department of Physics and Astronomy, 4242 TAMU, College Station, TX 77843, USA}
\affiliation{Texas A\&M Institute of Data Science
John R. Blocker Building, Suite 227
155 Ireland Street, TAMU 3156
College Station, TX 77843-3156}

\author[0000-0002-1210-2173]{Ulisses Braga-Neto}
\affiliation{Department of Electrical and Electronic Engineering, Texas A\&M University, College Station, TX, USA}

\author{Lifan Wang}
\affiliation{George P. and Cynthia Woods Mitchell Institute for Fundamental Physics \& Astronomy, \\
Texas A. \& M. University, Department of Physics and Astronomy, 4242 TAMU, College Station, TX 77843, USA}

\author{Daniel Kasen}
\affiliation{Astronomy Department and Theoretical Astrophysics Center, University of California, Berkeley, Berkeley, CA 94720, USA}

\author{Zhengwei Liu}
\affiliation{International Centre of Supernovae (ICESUN), Yunnan Key Laboratory of Supernova Research, Yunnan Observatories, Chinese Academy of Sciences (CAS), Kunming 650216, P.R. China}

\author[0000-0002-4460-0097]{Friedrich K.\ R{\"o}pke}
\affiliation{Heidelberger Institut f{\"u}r Theoretische Studien, Schloss-Wolfsbrunnenweg 35, 69118 Heidelberg, Germany}
\affiliation{Zentrum für Astronomie der Universit{\"a}t Heidelberg, Institut für Theoretische Astrophysik,
Philosophenweg 12, 69120 Heidelberg, Germany}
\affiliation{Zentrum für Astronomie der Universit{\"a}t Heidelberg, Astronomisches Rechen-Institut, M{\"o}nchhofstra{\ss}e 12--14, 69120 Heidelberg, Germany}

\author{Ming Zhong}
\affiliation{
Department of Mathematics
Philip Guthrie Hoffman Hall
3551 Cullen Blvd, Room 641, University of Houston, 
Houston, Texas 77204-3008
}

\author{David J. Jeffery}
\affiliation{Department of Physics \& Astronomy and Nevada Center for Astrophysics (NCfA), University of Nevada, Las Vegas, Nevada, U.S.A.}

\begin{abstract}

We present SEDONA-GesaRaT, a rapid code for supernova radiative transfer simulation developed based on the Monte-Carlo radiative transfer code SEDONA. 
We use a set of atomic physics neural networks (APNN), an artificial intelligence (AI) solver for the non-local thermodynamic equilibrium (NLTE) atomic physics level population calculation, which is trained and validated on 119 1-D type Ia supernova (SN Ia) radiative transfer simulation results showing great computation speed and accuracy. 
SEDONA-GesaRaT has been applied to the 3-D SN Ia explosion model N100 \citep{Seitenzahl2013N100} to perform a 3-D NLTE radiative transfer calculation. 
The spatially resolved linear polarization data cubes of the N100 model are successfully retrieved with a high signal-to-noise ratio using the integral-based technique (IBT) \citep{Chen2024IBT}. 
The overall computation cost of a 3-D NLTE spectropolarimetry simulation using SEDONA-GesaRaT is only $\sim$3000 core-hours, while the previous codes could only finish 1-D NLTE simulation, or 3-D local thermodynamic equilibrium (LTE) simulation, with similar computation resources. 
The excellent computing efficiency allows SEDONA-GesaRaT for future large-scale simulations that systematically study the internal structures of supernovae. 

\end{abstract}


\section{Introduction}\label{sec:intro}

Radiative transfer (RT) simulations are a critical tool in supernova (SN) research, providing the ability to synthesize observable predictions, including light-curves, spectra, and spectropolarimetry, based on the theoretical structures of supernovae (SNe). 
Many SN RT codes (e.g., TARDIS \citet{Kerzendorf2014Tardis}, CMFGEN, \citet{Hillier1998CMFGEN}) are based on a 1-D spherically symmetric approximation. 
Reasonable agreement between different RT programs has been shown \citep{Blondin2022Standart}, and spectral and light-curve modeling on both type Ia supernovae (SNe Ia) and core-collapse supernovae (CCSNe) has been successful in matching observations (e.g., \cite{Chen2024AIAI2,Dessart2014CMFIa,Vogl2020TypeII}). 
However, imaging of nearby SN remnants (e.g., \cite{Ferrazzoli2023Tycho,Temim2024CasA}) and spectropolarimetry observations of SNe (e.g., \cite{Cikota2019PolObs,Nagao2024IIPol}) suggest that both SNe Ia and CCSNe have complex 3-D structures, which are also revealed in multidimensional hydrodynamic simulations (e.g., \cite{Gamezo2004DDT,Kromer2013Hesma,Vartanyan2024TypeIIHyd}). 

Although several multidimensional RT codes have been developed (e.g., \cite{Kasen2006Sedona,Kromer2009ARTIS}) based on the Monte-Carlo method, 3-D time-dependent RT simulation remains a computationally challenging task due to the large number of Monte-Carlo quanta required to achieve high signal-to-noise ratio (S/N) results, and the large number of spatial grid points needed to solve for the plasma excitation and ionization state. 
To reduce the Monte-Carlo noise in the simulated spectropolarimetry result, \cite{Bulla2015VPack} proposed a virtual packet method. 
Based on the virtual packet method, \cite{Chen2024IBT} proposed an integral-based technique (IBT), which further increases the computational efficiency. 

Because a full non-local thermodynamic equilibrium (NLTE) calculation requires a detailed calculation of atomic transition rates to determine the statistical equilibrium of all the atomic levels, most 3-D RT simulations adopt a simple local thermodynamic equilibrium (LTE) or approximated NLTE to strike a balance between computational resources and physical fidelity. 
In some RT simulations, a level-merging method
\cite{Hoeflich2003Hydra,Anderson1985SuperLevel,Anderson1989StarNLTE} has been developed to reduce the number of atomic levels considered in NLTE for an approximate result. 
\cite{Boos2024OneDNLTE} proposed a data-driven method that synthesizes the 3-D NLTE RT spectra from the 3-D LTE simulation and the 1-D NLTE simulation using the extracted SN model along the line-of-sight. 

SEDONA \citep{Kasen2006Sedona} is a multi-dimensional Monte-Carlo RT code for SN spectral and light-curve simulation. 
Physics processes, including (1) gamma-ray emission from radioactive elements; (2) gamma-ray bound-free transition, Compton scattering, and photoelectric scattering; (3) optical photon bound-bound, bound-free, free-free transitions, are included in SEDONA simulation. 
An NLTE solver in SEDONA can be turned on for individual elements to determine the level populations in statistical equilibrium, but the computation cost is increased by a factor of $\sim$100 relative to LTE. 
This is a computation bottleneck for high-precision 3-D SN RT simulations. 
Inspired by the progress in machine learning accelerated computational fluid dynamics (e.g., \citep{Kochkov2021CFD}), we propose an artificial intelligence (AI) method to accelerate the NLTE calculation. 
In addition to the IBT algorithm \citep{Chen2024IBT} developed in our previous research, we present SEDONA-GesaRaT (the Gesamtkunstwerk of Radiative Transfer developed on SEDONA), an AI-accelerated RT program that is capable of 3-D NLTE SN RT simulations based on SEDONA\citep{Kasen2006Sedona}. 

Section \ref{sec:nlte} discusses the results of SNe Ia 1-D NLTE RT simulations, which serve as the training and testing data set for the AI.
Section \ref{sec:3d} tests the SEDONA-GesaRaT code on 3-D time-dependent RT simulations, and presents the first 3-D RT simulation result with both NLTE processes and spatially-resolved spectropolarimetry data cubes, using the SN Ia explosion model N100 \citep{Seitenzahl2013N100}. 
Section \ref{sec:conclusion} gives the conclusion. 

\section{NLTE Radiative Transfer}\label{sec:nlte}

\begin{figure*}[htb!]
    \includegraphics[width=0.245\textwidth]{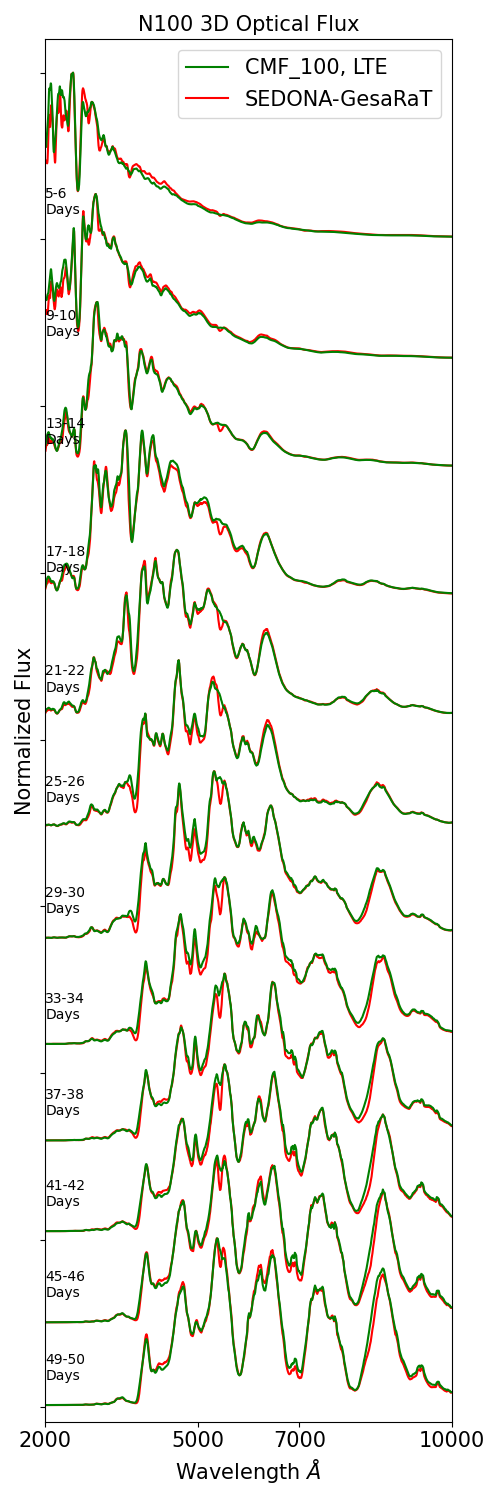}
    \includegraphics[width=0.245\textwidth]{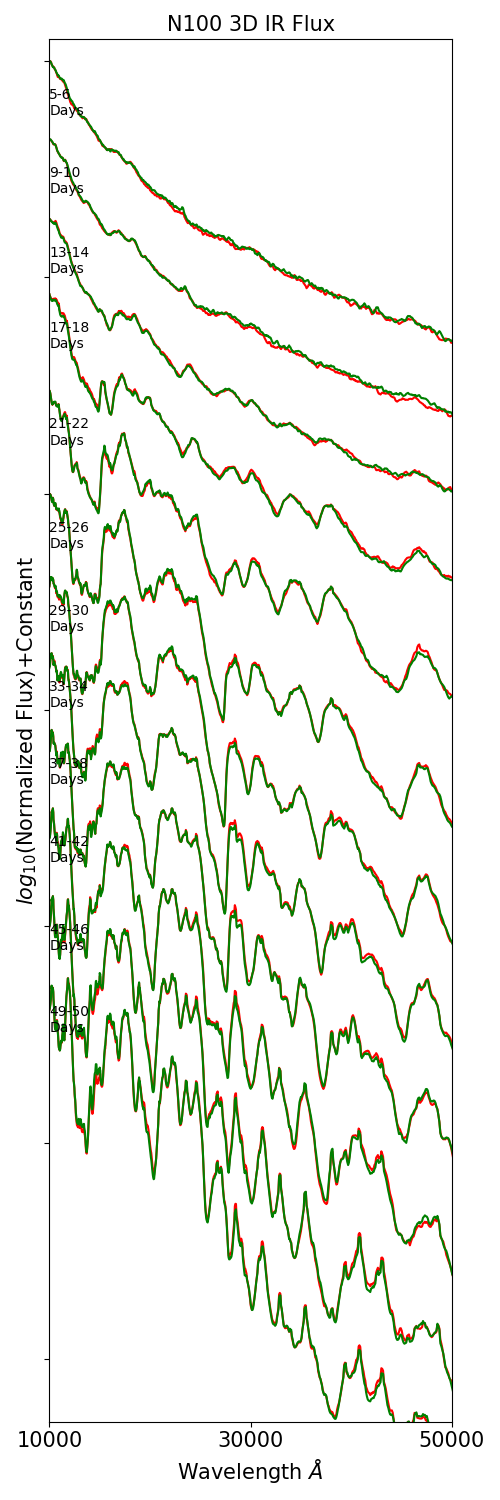}
    \includegraphics[width=0.245\textwidth]{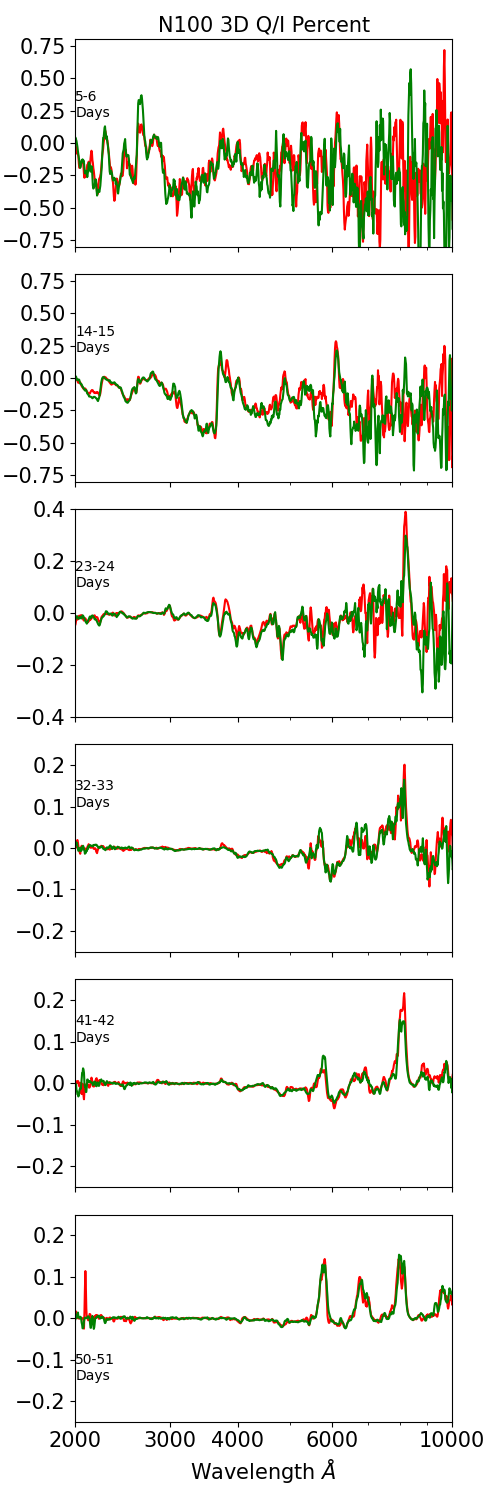}
    \includegraphics[width=0.245\textwidth]{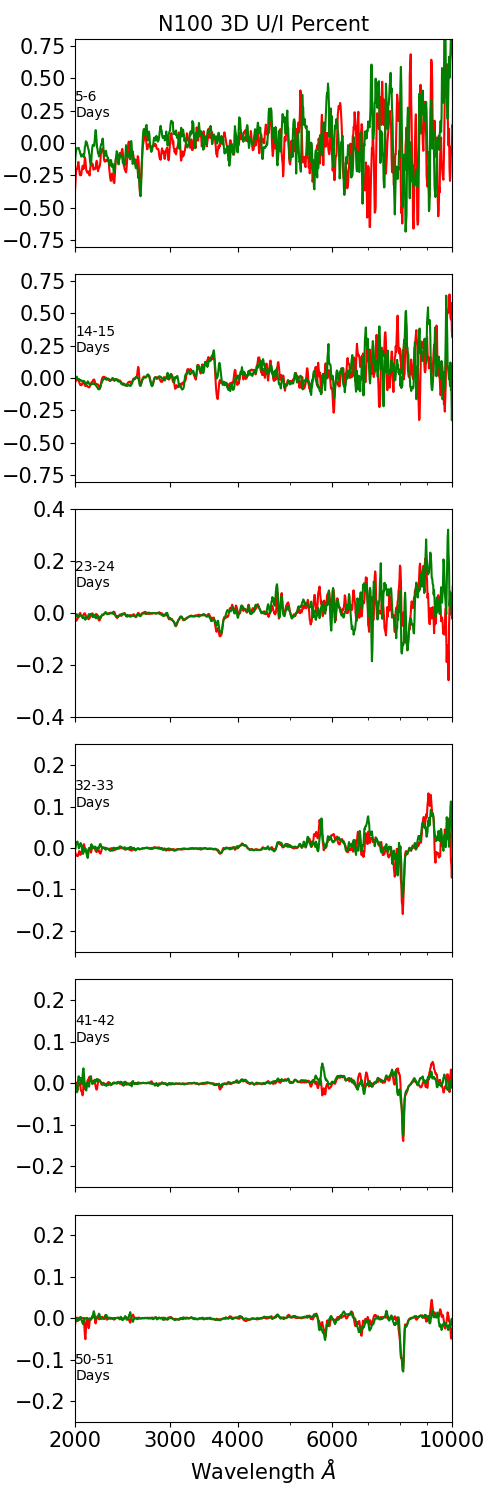}
    \caption{The spectropolarimetry time sequence of the 3-D N100 model at the viewing direction $\mu=0.5,\ \varphi=3.4455$. 
    The green line is calculated by SEDONA which use traditional method to solve the source function, while the red line is calculated by SEDONA-GesaRaT using APNN to solve the source function, both solutions use the CMF\_100 LTE recipe for atomic physics. 
    From left to right, the first panel shows the optical wavelength flux from $2000\ \rm{\AA}$ to $10000\ \rm{\AA}$, the second panel shows the infrared wavelength flux from $10000\ \rm{\AA}$ to $50000\ \rm{\AA}$, the third panel shows the linear polarization percentage $Q/I$, the fourth panel shows the linear polarization percentage $U/I$. 
    The $x$ axes in the third and the fourth panels are in logarithmic scale for better presentation. }
    \label{fig:ltecomp}
\end{figure*}

We downloaded the 119 SN Ia ejecta structures from the Heidelberg Supernova Model Archive (HESMA) \footnote{\href{https://hesma.h-its.org/}{https://hesma.h-its.org/}} \citep{Kromer2017Hesma}, then used SEDONA \citep{Kasen2006Sedona} to conduct a 1-D time-dependent RT simulation of all the ejecta structures. 
Lists of the ejecta structures are shown in Appendix \ref{sec:apnn}. 
In the 1-D time-dependent RT simulation, we use two sets of atomic data libraries: CMF\_All and CMF\_100. 
The atomic data library CMF\_100 is reduced from the CMFGEN \footnote{\href{https://sites.pitt.edu/~hillier/web/CMFGEN.htm}{https://sites.pitt.edu/$\sim$hillier/web/CMFGEN.htm}}\citep{Hillier1998CMFGEN} database to limit the data for each ion to no more than the 100 lowest lying levels; the total number of levels is 28,105, and the total number of transitions is 274,128. 
CMF\_All includes all of the atomic levels and transitions from the CMFGEN database, but the Si, S, and Ca elements are kept the same as CMF\_100 in order to control the NLTE atomic physics computation time; this amounts to, for all ionic species, 497,424 levels and 35,344,426 bound-bound transitions. 
The forbidden lines (e.g., Ca II 7291$\rm{\AA}$, Ca II 7324$\rm{\AA}$) are not included in the two atomic data libraries. 

The HESMA ejecta models are specified at 100 seconds after the explosion. 
Before the RT simulation, the ejecta structure is remapped to day 1 after the explosion, assuming adiabatic and homologous expansion, and the energies from the decay of radioactive elements before day 1 are converted to local thermal energies, which act as the initial plasma temperatures in the following RT calculations. 
In SEDONA 1D time-dependent RT calculations, the energy deposition of radioactive isotopes $^{56}$Co and $^{56}$Ni is calculated by transporting gamma-ray energy packets and determining the energy they deposit in the ejecta due to Compton scattering and photoionization. 
The radioactive energy deposited in the ejecta is re-emitted as thermal (UV/optical/infrared) photons. 
The included interaction physics are Thomson scattering, with a comoving frame extinction coefficient: 
\begin{equation}
    \bar{k}_e=\sigma_T ~n_e\ , 
\end{equation}
where $\sigma_T$ is the Thomson scattering cross section, $n_e$ is the plasma free electron number density. 
The bound-bound line opacity is calculated under the Sobolev approximation and using an expansion opacity formalism \citep{Karp1977LineExpansion}. 
Specifically, the individual Sobolev line optical depth is calculated as
\begin{equation}
    \tau_{\rm Sob}=\frac{h}{4\pi}\left(N_{l}B_{lu}-N_{u}B_{ul}\right)t_{\rm{exp}}c\ , 
\end{equation}
where $h$ is the Planck constant, $B_{lu}$ and $B_{ul}$ are Einstein coefficients, $N_{l}$ and $N_{u}$ are the level population number density of the lower level ($l$) and the upper level ($u$), $t_{\rm exp}$ is the time after the SN explosion, $c$ is the speed of light. 
The probability that a photon interacts with a line when coming into resonance with it is
\begin{equation}
    \tau_{\rm eff}=1-e^{-\tau_{\rm Sob}}\ . 
\end{equation}
In the expansion opacity formalism, the total bound-bound extinction coefficient is calculated by summing all lines within a given frequency bin 
\begin{equation}
    \bar{k}_{bb}(\bar{\nu},\bar{\nu}+\Delta\bar{\nu})=\frac{(\bar{\nu}+0.5\Delta \bar{\nu})\sum_{i}\tau_{eff,i}}{\Delta \bar{\nu}\ t_{\rm exp}\ c}\ , 
\end{equation}
where $\Delta \bar{\nu}$ is the frequency bin width in the comoving frame, the summation is over all the spectral lines in the frequency bin. 
The bound-bound extinction coefficient is summed with the bound-free extinction coefficient and the free-free extinction coefficient, which are calculated in the same frequency bin, to form the total extinction coefficient $\bar{k}$. 
The total emission coefficient $\bar{j}$ from bound-bound, bound-free, and free-free transitions is also calculated in each frequency bin. 
Note that in the expansion opacity formalism, the individual line profiles are not resolved. 
This approximation should be reasonable with the bin only including optically thin lines ($\tau_{\rm sob} < 1$). 
If the bin includes multiple optically thick lines, however, the $\bar{J}_{\nu}$ in each line may be less reliable. 
Further NLTE studies are needed to compare the differences between resolved line transport and the expansion opacity formalism.

We evolve the RT simulation until 60 days after the explosion. 
Before 14.78 days, the simulation time step is logarithmic with $\Delta t_{\rm{s}}/t_{\rm{s}}=0.02$. 
After 14.78 days, the simulation time step is $\Delta t_{\rm{s}} = 0.2 $ days. 
In each time step, $5\times 10^{4}$ energy packets are created. 
The frequency grid to calculate $\bar{J_{\nu}}$, $\bar{k}$, $\bar{j}$ is a logarithmic grid from $5\times 10^{13}$ Hz to $5\times 10^{15}$ (600 to 60000 $\rm{\AA}$) with $\Delta \nu / \nu =0.002$; the total number of frequency grid points is 2305. 
After each time step and at each radial point, we record the elemental abundances, plasma density, mean intensity of the radiation field, extinction coefficients, and emissivities. 

For each ejecta model, we run four RT calculations using different atomic physics libraries and approximations: 

\begin{enumerate}
    \item Using the CMF\_100 atomic library, with all elements in LTE  (CMF\_100 LTE). 
    \item Using the CMF\_All atomic library, with calcium calculated in NLTE and all other elements in LTE (CMF\_All Ca NLTE). 
    \item Using the CMF\_All atomic library, with calcium and silicon calculated in NLTE and all other elements in LTE (CMF\_All Si, Ca NLTE). 
    \item Using the CMF\_All atomic library, with calcium, sulfur, and silicon in NLTE, and all other elements in LTE (CMF\_All Si, S, Ca NLTE). 
\end{enumerate}

One of the most expensive aspects of the simulations is calculating the extinction coefficients and emissivities, as this requires a solution of the atomic level populations (involving a costly solution to the rate equations in the NLTE case) and a subsequent calculation of all relevant radiative processes at each frequency point. 
To accelerate this aspect of the code, we recorded the results of this calculation in every radial zone and at every time step for each of the 119 ejecta models. 
We used this data to train a set of atomic physics neural networks (APNNs) that were able to quickly map a given zone density and composition to the resulting vectors of the extinction coefficient and emissivity as a function of frequency. 
Because our training set covered the range of physical conditions relevant for SNe Ia, the APNN should be applicable for general usage for SNe Ia modeling. 
The details of the neural network training are discussed in Appendix \ref{sec:apnn}, and the testing results of APNN are illustrated in Appendix \ref{sec:1d}. 
\section{Test on 3-D Models}\label{sec:3d}

In this section, we use the SN Ia explosion model N100 \citep{Seitenzahl2013N100} to conduct a 3-D time-dependent RT simulation. 
For this purpose, we adopt the end stage of the hydrodynamic evolution of the N100 model together with the nucleosynthetic postprocessing result, and map the ejecta structure on a $50\times50\times50$ Cartesian spatial grid, and the maximum ejecta velocity is set to 28600 km/s. 

SEDONA-GesaRaT uses the integral-based technique (IBT) \citep{Chen2024IBT}. 
IBT retrieves the spectropolarimetry time sequence from a 3-D Monte Carlo RT simulation of SN at a specified viewing direction. 
Comparing to the default spectral retrieval method in SEDONA (direct-counting technique, DCT), IBT increases the Monte Carlo signal-to-noise ratio (S/N) by a factor of $\sim$30 while only increasing the computation time by $\sim$30\%, allowing SEDONA-GesaRaT to retrieve high S/N spectropolarimetry with limited computation time. 

\subsection{LTE Results Comparison}


To test the performance of APNN in 3-D time-dependent RT simulations, we adopt the CMF\_100 LTE atomic physics parameters to perform RT simulations on the N100 model using SEDONA and SEDONA-GesaRaT. 
The complete atomic physics library and the NLTE calculation are not adopted in this comparative simulation due to the excessive computation time using the traditional level population solver as shown in Table \ref{tab:apnntime}. 
Both simulations start at 1 day after the explosion, end at 60 days after the explosion, and the number of energy packets per step is $4\times 10^{6}$. 
Before 14.78 days, the simulation time step is logarithmic with $\Delta t_{s}/t_{s}=0.02$, after 14.78 days, the simulation time step is $\Delta t_{s}=0.2$ days. 
The total simulation time is $\sim$73.3 hours for SEDONA and $\sim$56.3 hours for SEDONA-GesaRaT, measured on a 48-core computing node on TAMU HPRC Grace supercomputer \footnote{\href{https://hprc.tamu.edu/kb/User-Guides/Grace/}{https://hprc.tamu.edu/kb/User-Guides/Grace/}}, which is mounted with 2 chips of Intel Xeon Gold 6248R CPU. 

Figure \ref{fig:ltecomp} compares the spectropolarimetric time sequence from 2000 $\rm{\AA}$ to 50000 $\rm{\AA}$ between SEDONA and SEDONA-GesaRaT. 
We notice that the SEDONA-GesaRaT results are consistent with the SEDONA simulation results on spectral flux between 2000 $\rm{\AA}$ and 50000 $\rm{\AA}$ from 5 days to 50 days after the explosion. 
The linear polarization results from SEDONA-GesaRaT are also consistent with the results from SEDONA between 2000 $\rm{\AA}$ and 10000 $\rm{\AA}$, subject to the Monte-Carlo noise. 
However, we notice the opacity of the Si II 5640$\rm{\AA}$ spectral line is overestimated in SEDONA-GesaRaT, which causes an extra absorption feature around 17 days after the explosion. 
A similar issue could also be found on the Ca II infrared triplet spectral lines around 41 days after the explosion. 
This error indicates that the AI prediction performance could be further improved by refining the neural network structure and including more diverse SN Ia ejecta models in the training data set. 
On the other hand, the tests of SEDONA-GesaRaT on 1-D time dependent simulations shown in Figure \ref{fig:1dspec} did not show further decreases of accuracy with the inclusion of more complex atomic physics libraries and more elements for NLTE calculations. 
We conclude the current configuration of APNN is appropriate for most of 3-D NLTE RT simulations and for the exploration of the NLTE effect on the spectropolarimetry signals. 

\subsection{NLTE Results}

\begin{figure*}
    \includegraphics[width=\textwidth]{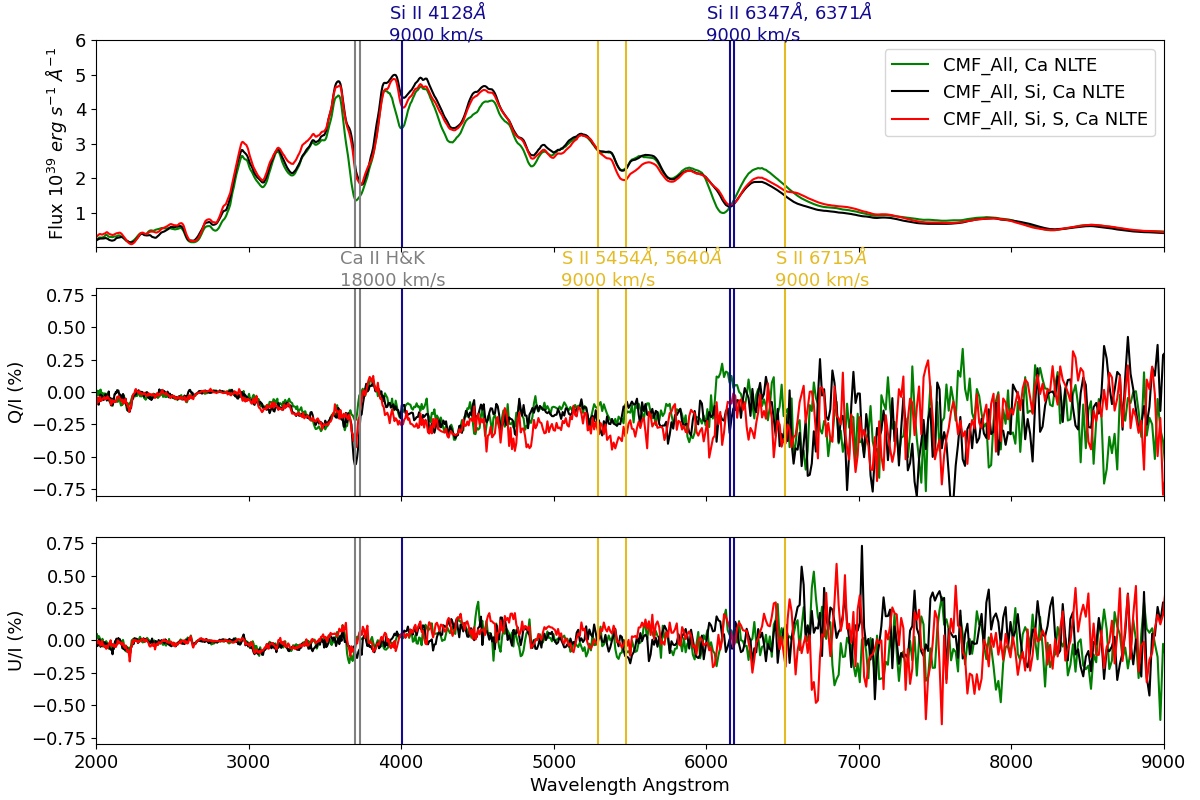}
    \caption{The spectropolarimetry of the N100 model between 17 and 18 days after the explosion, in the viewing direction $\mu=0.5$, $\varphi=3.4455$. Upper panel is the spectral flux, middle panel is the linear polarization percentage $Q/I$, lower panel is the linear polarization percentage $U/I$. The identified spectral lines are marked with vertical lines and labeled with the element names and the blueshift velocities. }
    \label{fig:n3d17}
\end{figure*}

In this section, we use SEDONA-GesaRaT to calculate the spectropolarimetric time series of the 3-D N100 model using Si, S, and/or Ca in NLTE. 
Figure \ref{fig:n3d17} shows the flux and linear spectropolarimetry of the N100 model between 17 and 18 days after the explosion in the viewing direction $\mu=0.5$, $\varphi=3.4455$. 
Comparing the spectropolarimetry results under different atomic physics recipes, we notice that: (1) when treating sulfur in NLTE the emission feature of the S II 6715$\rm{\AA}$ line is enhanced, and the absorption features of S II 5454$\rm{\AA}$ and 5640$\rm{\AA}$ are enhanced; (2) when treating silicon in NLTE, the P-Cygni profile of Si II 6347$\rm{\AA}$ and 6371$\rm{\AA}$ and the absorption feature of Si II 4128$\rm{\AA}$ lines are enhanced, and the linear polarization signal at 6100$\rm{\AA}$ is suppressed. 

\begin{figure}
    \centering
    \plotone{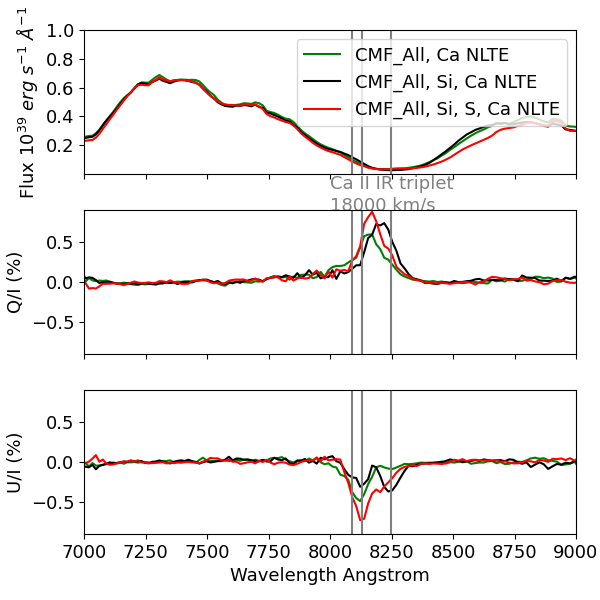}
    \caption{Same as Figure \ref{fig:n3d17}, but between 45 and 46 days after the explosion and focused on 7000--9000 $\rm{\AA}$. The Ca II infrared triplet lines are marked with gray vertical lines. }
    \label{fig:n3d45}
\end{figure}

High polarization usually occurs at the absorption feature P-Cygni lines, as shown in Figure \ref{fig:n3d17} because unpolarized light from the photosphere is scattered out of the line-of-sight by the spectral line \citep{Jeffrey1991SpecPol}. 
We note that treating an element in NLTE may have indirect effects on the spectral features of other elements. 
For example, Figure \ref{fig:n3d17} shows that putting silicon in NLTE modifies the spectropolarimetric feature near $4000\rm{\AA}$, which is primarily due to the Ca II H\&K lines but is affected by blended SiII lines. The silicon also generally increases the flux between 4000$\rm{\AA}$ and 5000$\rm{\AA}$. 
Moreover, the linear spectropolarimetry of the Ca II infrared triplet at 45 days after the explosion also is changed by the inclusion of Si and S elements in NLTE (Figure \ref{fig:n3d45}). 

\subsection{Resolved Image}\label{sec:resoimg}

\begin{figure*}[htb!]
    \centering
    \includegraphics[width=\textwidth]{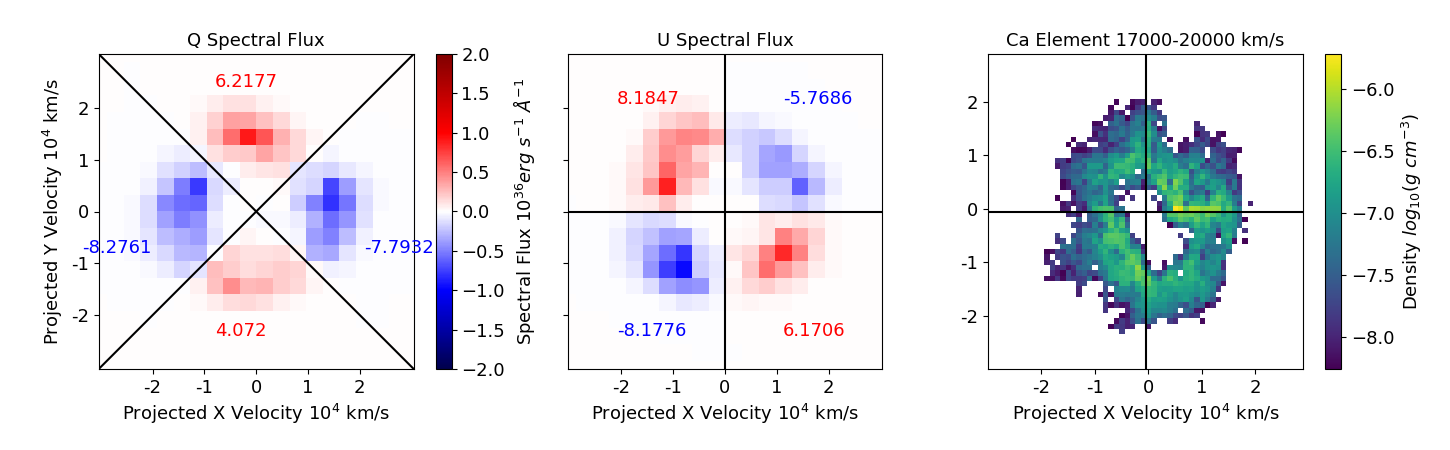}
    \includegraphics[width=\textwidth]{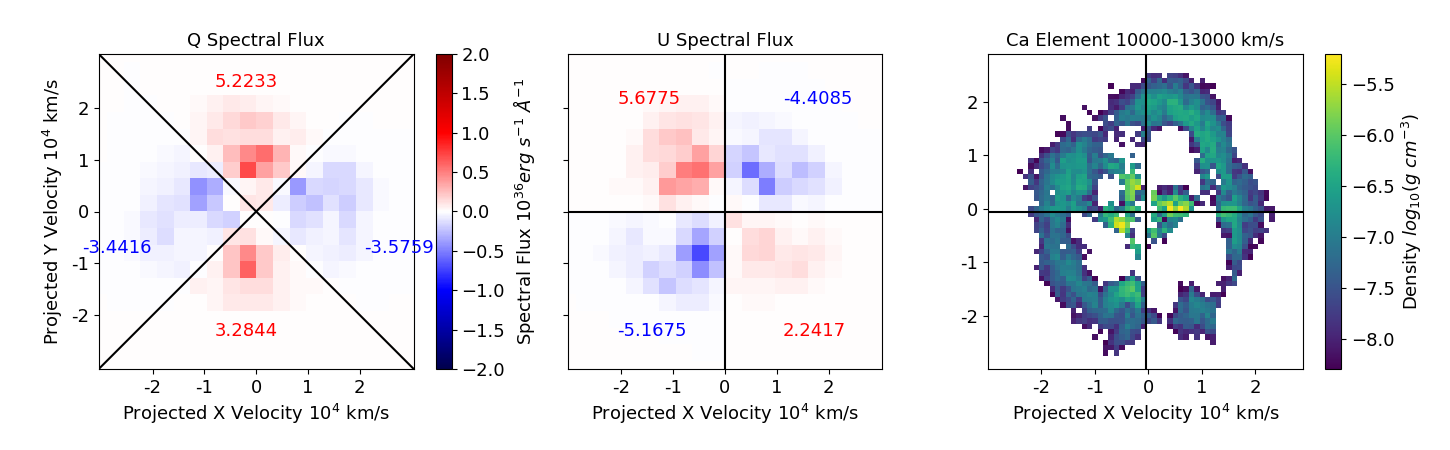}
    \caption{Left and middle panel show the linear polarization maps of the Q and U components at 17-18 days after the explosion. 
    Upper panel is at the wavelength 3697 $\rm{\AA}$ and lower panel is at the wavelength 3800 $\rm{\AA}$. 
    The maps are split into four regions following the value symbol. 
    The sum of polarization value in each region are shown with numbers in the unit of $10^{36} {\rm erg\ s^{-1}\ \AA^{-1}}$. 
    Upper right panel shows the Ca element number density in a section of the N100 model in the line-of-sight velocity 17000--20000 km/s. 
    Lower right panel shows the Ca element number density in a section of the N100 model in the line-of-sight velocity 10000--13000 km/s. }
    \label{fig:pol1394}
\end{figure*}

SEDONA-GesaRaT can also simulate spatially resolved spectropolarimetry data cubes based on the IBT method \citep{Chen2024IBT}. 
Using the N100 model, the spectropolarimetry data cube is calculated along the viewing direction $\mu=0.5,\ \varphi=3.4455$. 
The number of energy packets is $4\times 10^{6}$ per time step, which is higher than that in Section \ref{sec:nlte}, so as to reduce the Monte-Carlo noise in the resolved image. 
The total simulation time is increased to 63 hours on a 48-core computation node at the TAMU HPRC Grace supercomputer, giving a total simulation cost of 3024 core-hours. 
The observer image grid is defined on the SN projected image from $-32000\times t_{\rm exp}\ {\rm km/s}$ to $32000\times t_{\rm exp}\ {\rm km/s}$, where $t_{\rm exp}$ is the time after the SN explosion, and the grid number of X and Y axes is both 20. 

The linear polarization image between 17 and 18 days after the explosion at 3697 $\rm{\AA}$ and at 3800 $\rm{\AA}$ are shown in Figure \ref{fig:pol1394}. 
Note that the two selected wavelength values are broadly correlated with the Ca II H\&K spectral lines with line-of-sight velocities 17000--20000 km/s and 10000--13000 km/s, and the Ca element density on sections of the N100 model in the corresponding velocity ranges is also shown in Figure \ref{fig:pol1394}. 
We notice that the Q and U polarization maps can be divided into four quadrants based on the positivity of signals, which is in agreement with Figure 12 of \cite{Bulla2015VPack}. 
Moreover, the linear polarization intensity is broadly overlapped with the Ca element density map in the corresponding line-of-sight velocity region. 
Specifically, the Ca element density is zero at the center of the 17000-20000 km/s line-of-sight section, which is in agreement with the zero polarization intensity at the center of the 3697 $\rm{\AA}$ polarization map; the Ca element density reaches $\sim 10^{-5.5}\ {\rm g\ cm^{-3}}$ at the center of the 10000--13000 km/s line-of-sight section, which is in agreement with the non-zero polarization intensity at the center of the 3800 $\rm{\AA}$ polarization map; the lower-right part of the 10000--13000 km/s line-of-sight section shows low Ca element density, which is in agreement with weak U linear polarization in the same region. 

\section{Conclusion}\label{sec:conclusion}

We presented SEDONA-GesaRaT, an accelerated code for SN RT simulation based on SEDONA\citep{Kasen2006Sedona}. 
SEDONA-GesaRaT is capable of 3-D simulations with the inclusion of NLTE atomic physics processes, realized by APNN trained on 119 1-D SN Ia RT simulation results.  
Using the IBT spectral synthesis algorithm \citep{Chen2024IBT}, SEDONA-GesaRaT is capable of retrieving spatially resolved flux and linear polarization data cubes in a specified viewing direction. 
The tests of SEDONA-GesaRaT on 1-D RT simulations demonstrate that the APNN is capable of emulating the complicated NLTE processes in the SN Ia plasma, and reducing the computation time from $\sim$50 core-seconds to 0.17 core-seconds. 
The tests on 3-D RT simulations using the N100 SN Ia ejecta model under the LTE approximation show that SEDONA-GesaRaT could calculate the spectropolarimetry time sequence, and the result is in agreement with SEDONA. 
Finally, we have conducted the 3-D RT simulation of the N100 SN Ia ejecta model, with the inclusion of Si, S, and Ca for NLTE processes, and present the spatially resolved spectropolarimetry data cube as a time sequence. 

Minor inaccuracy of APNN is observed on the S II 5640$\rm{\AA}$ spectral line in 3-D LTE simulations. 
This inaccuracy could be alleviated by including the 1-D RT simulation results with different numbers of Monte-Carlo quanta during the generation of the training and validating data, and refining the neural network structures. 
Although the 1-D RT simulated spectral time sequence accuracy is not reduced, the mean squared error (MSE) loss measured on the validating data set increases with the inclusion of more elements for the NLTE processes in APNN training. 
We also notice that the inclusion of more elements (e.g., Si, S) in NLTE could result in the change of other elements' (e.g., Ca) spectopolarimetric line shapes. 
These phenomena suggest that the further upgrade of SEDONA-GesaRaT will require a more physically complete NLTE solution, which depends on both the accuracy of atomic physics data and the stability of the NLTE solver. 
A further comparison of atomic physics data libraries (e.g., \cite{Kurucz1995atomdata,Del2021Chianti}) and NLTE solvers (e.g., \cite{Mazzali1993DiluteLte,Dessart2014CMFIa}) is necessary. 
Furthermore, with the current APNN being trained on 1-D SN Ia ejecta models, the application of SEDONA-GesaRaT on core-collapse SNe will require new training data with necessary elemental abundances (e.g., hydrogen and helium elements). 

Our previous researches have proposed an artificial-intelligence assisted inversion (AIAI) method to estimate the SN Ia ejecta structure from the observed spectra, which has been successfully applied onto $\sim$100 observed SNe Ia \cite{Chen2020AIAI,Chen2024AIAI2}. 
While the AIAI method relies on a large-scale RT simulation to create a SN Ia model and spectral libraries, the previous researches are limited to simple 1-D time-dependent RT simulations due to computation resources. 
The newly-developed SEDONA-GesaRaT provides an efficient solution for the complex 3-D NLTE RT simulations, making it ideal for the further investigation of SN Ia ejecta structures in 3-D, and the modelling of SN Ia spectropolarimetry observations (e.g., \cite{Cikota2019PolObs}). 

\clearpage

\begin{acknowledgments}

X.C. is supported by the National Science Foundation under award NSF-AST 1817099, and is supported by Texas A\&M University Institute of Data Science (TAMIDS). 
The work of U. B-N. is supported by the National Science Foundation under award CCF-2225507. 
The work of Z.W.L.\ is supported by the Strategic Priority Research Program of the Chinese Academy of Sciences (grant Nos. XDB1160303, XDB1160000), the National Natural Science Foundation of China (NSFC, Nos.\ 12288102, 12090040/1, 11873016), the National Key R\&D Program of China (Nos.\ 2021YFA1600401 and 2021YFA1600400), the International Centre of Supernovae (ICESUN), Yunnan Key Laboratory of Supernova Research (No. 202302AN360001), and the Yunnan Fundamental Research Projects (grant Nos.\ 202201BC070003, 202001AW070007).
The work of F.K.R.\  is supported by the Klaus Tschira Foundation, by the Deutsche Forschungsgemeinschaft (DFG, German Research Foundation) -- RO 3676/7-1, project number 537700965, and by the European Union (ERC, ExCEED, project number 101096243). Views and opinions expressed are, however, those of the authors only and do not necessarily reflect those of the European Union or the European Research Council Executive Agency. Neither the European Union nor the granting authority can be held responsible for them. 

Portions of this research were conducted with the advanced computing resources provided by Texas A\&M High Performance Research Computing. 
This work used FASTER super computer at TAMU HPRC through allocation PHY240215 from the Advanced Cyberinfrastructure Coordination Ecosystem: Services \& Support (ACCESS) program, which is supported by National Science Foundation grants 2138259, 2138286, 2138307, 2137603, and 2138296 \citep{boerner2023access}. 
This work made use of the Heidelberg Supernova Model Archive (HESMA), https://hesma.h-its.org \citep{Kromer2013Hesma}. 

The authors would like to thank Prof. Edward Baron from The Planetary Science Institute, Prof. J. C. Wheeler from University of Texas at Austin for supportive discussions. 
\end{acknowledgments}

\software{SEDONA\citep{Kasen2006Sedona},pytorch\citep{paszke2017pytorch}}


\appendix

\section{Atomic Physics Neural Network}\label{sec:apnn}

Using the RT simulation results discussed in Section \ref{sec:nlte}, we trained an atomic physics neural network (APNN), which uses the elemental abundances, plasma density, and $\bar{J_{\nu}}$ as input to predict $\bar{j}$, $\bar{k}$, and $\bar{\sigma}$. 
Throughout the SN explosion process, the physical quantities vary across multiple orders of magnitude; we therefore design a normalization scheme in order to keep the calculation within the APNN dynamic range. 
The new APNN inputs are: 

\begin{itemize}
    \item The scalar $\rho_{\rm new}$, which is the rescaled plasma density $\rho_{\rm new}=\left(log_{10}(\bar{\rho})+23\right)/4$ . 
    \item The 2305-element array $\bar{J}_{\rm nu}$, which is $\bar{J}_{\nu}$ divided by its maximum: $\bar{J}_{\rm nu}=\bar{J_{\nu}}/ {\rm Max}({\bar{J_{\nu}}})$ . 
    \item The scalar $\bar{E}$, which is the rescaled radiation energy integrated from $\bar{J}_{\nu}$: $\bar{E}=\left(log_{10}(\sum_{i=0}^{2305}\bar{J}_{\nu}\Delta\nu_{i})-10.5\right)/3.5$ . 
    \item The 28-element array $N_{\rm elem}$, which is the elemental abundances from hydrogen to nickel. 
\end{itemize}

The new APNN outputs are: 

\begin{itemize}
    \item The scalar $k_{\rm hi}$, which is the rescaled maximum of the absorption coefficient value: $k_{\rm hi}=\left(log_{10}({\rm Max}(\bar{k}))-2.5\right)/2.5$ . 
    \item The scalar $k_{\rm lo}$, which is the rescaled minimum of the absorption coefficient value: $k_{\rm lo}=\left(log_{10}({\rm Min}(\bar{k}))+3\right)/3$ . 
    \item The 2305-element array $k_{\rm nu}$, which is the rescaled absorption coefficient $k_{nu}=\frac{(log_{10}(\bar{k})-log_{10}({\rm Min}(\bar{k})))}{(log_{10}({\rm Max}(\bar{k}))-log_{10}({\rm Min}(\bar{k})))}$ . 
    \item The scalar $j_{\rm hi}$, which is the rescaled maximum of the emission coefficient value: $j_{\rm hi}=\left(log_{10}({\rm Max}(\bar{j}))+13.5\right)/1.5$ . 
    \item The scalar $j_{lo}$, which is the rescaled minimum of the emission coefficient value: $j_{\rm lo}=\left(log_{10}({\rm Min}(\bar{j}))+19\right)/3$ . 
    \item The 2305-element array $j_{\rm nu}$, which is the rescaled emission coefficient $j_{\rm nu}=\frac{(log_{10}(\bar{j})-log_{10}({\rm Min}(\bar{j})))}{(log_{10}({\rm Max}(\bar{j}))-log_{10}({\rm Min}(\bar{j})))}$ . 
    \item The scalar $\sigma_{e}$, which is rescaled electron scattering opacity $\sigma_{e}=(log_{10}(\bar{k}_e)+3)/3$ . 
\end{itemize}

\begin{figure}[htb!]
    \centering
    \includegraphics[width=0.7\linewidth]{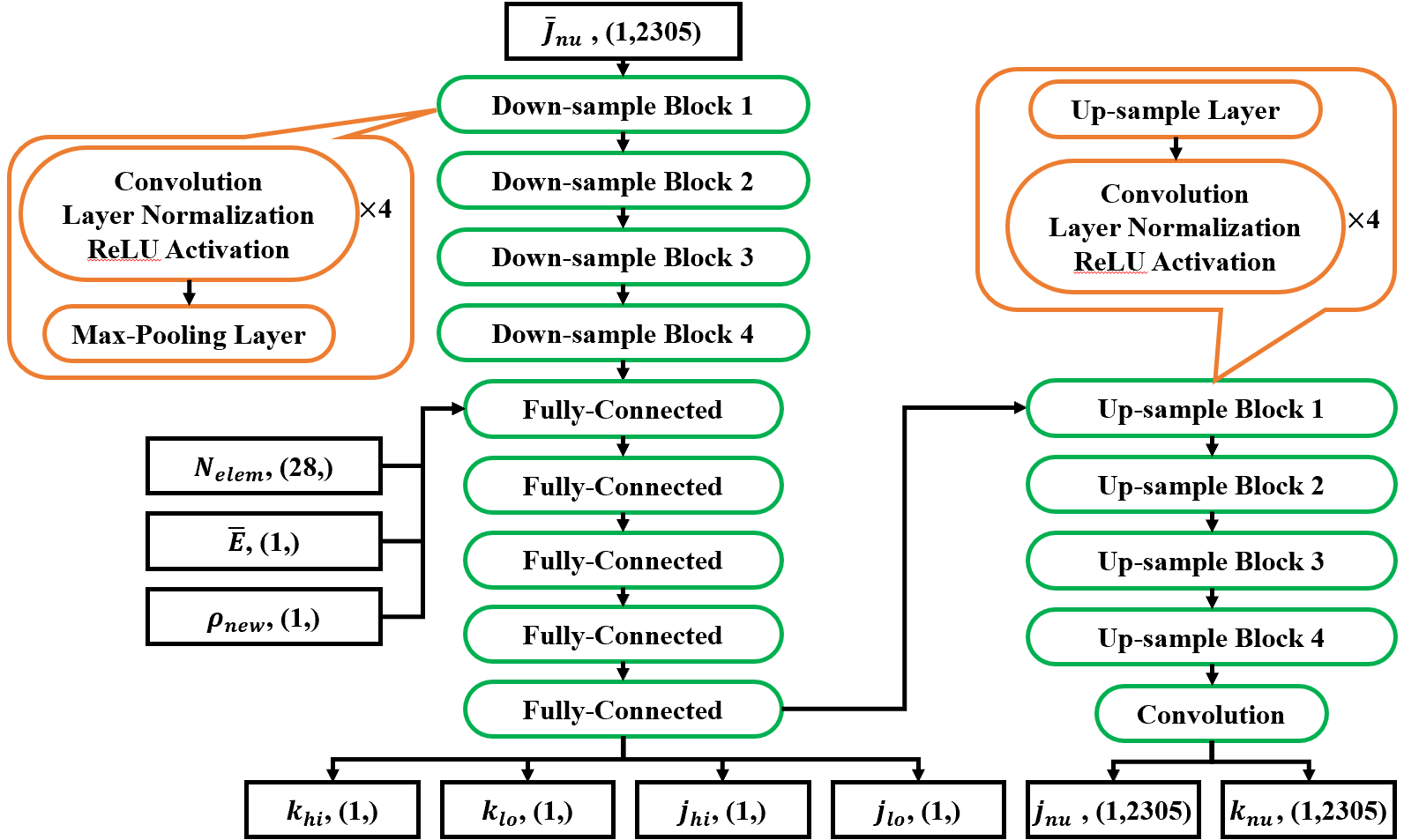}
    \caption{The illustrative neural network structure for APNN. }
    \label{fig:netillu}
\end{figure}

We adopt a convolution neural network (CNN) structure in APNN, with 33 convolution layers, 5 fully-connected layers, and 13,480,395 trainable parameters in total. 
In the APNN, the input $\bar{J}_{\rm nu}$ is firstly encoded with four down-sampling blocks, each block consists of four sub-blocks of convolution, layer normalization, and ReLU activation layers, and a max-pooling layer at the end of the block. 
The output of the fourth down-sampling block is concatenated with the input $N_{\rm elem}$, $\bar{E}$, and $\rho_{\rm new}$, then input into a stack of five fully-connected layers to predict $j_{\rm hi}$, $j_{\rm lo}$, $k_{\rm hi}$, and $k_{\rm lo}$. 
The output of the fourth fully-connected layer is input into a decoder network which consists of four up-sampling blocks and a convolution layer, to predict $j_{\rm nu}$ and $k_{\rm nu}$. 
In Figure \ref{fig:netillu}, we illustrate the overall structure of the APNN. 
The detailed neural network structure is shown in \href{https://geronimochen.github.io/images/APNN.png}{https://geronimochen.github.io/images/APNN.png}, and all the hyper-parameters can be seen in the online picture with magnification. 
Note that the APNN is defined on the frequency grid from $5\times 10^{13}$ Hz to $5\times 10^{15}$ Hz, adopting the APNN to other researches with different frequency grid will require extra interpolating effort. 

Among the 119 SNe Ia ejecta structures and the corresponding 1-D RT simulation results, 80 of them listed in Table \ref{tab:trainset} are in the training data set, 20 of them listed in Table \ref{tab:valiset} are in the validating data set, and 19 of them listed in Table \ref{tab:testset} are in the testing data set. 
The training data set is used in the APNN training process to update the trainable parameters; the validating data set is not used to update the trainable parameters, but to monitor the APNN prediction accuracy during the training process; the testing data set is only used to run 1-D RT simulation to compare the results between APNN and traditional methods. 

\begin{table}[htb!]
    \centering
    \begin{tabular}{c|c||c|c}
        \hline
        Model Name & Number of Shells & Model Name & Number of Shells \\
        \hline
        \hline
        gcd\_2021\_r10\_d2.6 & 83 & gcd\_2021\_r10\_d1.0 & 83 \\
        ddt\_2013\_n1600 & 95 & def\_2021\_r60\_d2.6\_z & 75 \\
        merger\_2010\_09\_09 & 93 & doubledet\_2012\_csdd-s & 60 \\
        ddt\_2013\_n1600c & 97 & ddt\_2014\_rpc32\_ddt8\_n100 & 98 \\
        det\_2015\_one13e7 & 77 & def\_2021\_r60\_d2.0\_z\_rot1 & 83 \\
        doubledet\_2021\_m0810r\_1 & 65 & ddt\_2013\_n40 & 91 \\
        doubledet\_2020\_2a\_21 & 77 & def\_2014\_n1600def & 98 \\
        def\_2014\_n1600cdef & 92 & def\_2014\_n20def & 98 \\
        ddt\_2013\_n100h & 91 & def\_2021\_r57\_d3.0\_z & 82 \\
        def\_2021\_r163\_d2.0\_z & 75 & def\_2021\_r143\_d3.0\_z & 77 \\
        ddt\_2014\_c50\_ddt8\_n100 & 98 & gcd\_2021\_r65\_d2.0 & 83 \\
        det\_2015\_one15e7 & 73 & def\_2021\_r60\_d2.6\_0.1z & 80 \\
        merger\_2013\_09\_076\_z1 & 73 & doubledet\_2021\_m0903\_1 & 91 \\
        doubledet\_2020\_2a & 77 & def\_2014\_n150def & 95 \\
        hedet\_2012\_hed-l & 61 & def\_2021\_r10\_d3.0\_z & 93 \\
        def\_2021\_r82\_d1.0\_z & 90 & gcd\_2016\_gcd200 & 67 \\
        ddt\_2013\_n100 & 93 & doubledet\_2021\_m0905\_1 & 74 \\
        def\_2021\_r206\_d1.0\_z & 72 & doubledet\_2020\_1a & 78 \\
        doubledet\_2021\_m0910\_1 & 59 & def\_2021\_n5\_d2.6\_z & 88 \\
        gcd\_2021\_r60\_d2.6 & 83 & gcd\_2021\_r150\_d2.6 & 83 \\
        def\_2014\_n300cdef & 93 & def\_2014\_n5def & 89 \\
        ddt\_2013\_n100l & 92 & def\_2014\_n1def & 84 \\
        ddt\_2013\_n150 & 92 & def\_2021\_r150\_d2.6\_z & 71 \\
        def\_2014\_n100def & 97 & det\_2010\_0.97 & 74 \\
        det\_2015\_co15e7 & 82 & ddt\_2013\_n200 & 94 \\
        doubledet\_2021\_m1005\_1 & 75 & doubledet\_2020\_2a\_79 & 76 \\
        doubledet\_2020\_3a & 59 & def\_2021\_r10\_d6.0\_z & 95 \\
        def\_2015\_n5\_hybrid & 34 & doubledet\_2020\_2b & 78 \\
        doubledet\_2020\_2a\_36 & 76 & def\_2014\_n100ldef & 97 \\
        doubledet\_2021\_m0805\_1 & 73 & merger\_2012\_11\_09 & 67 \\
        ddt\_2013\_n20 & 86 & det\_2015\_one20e7 & 73 \\
        hedet\_2012\_hed-s & 61 & merger\_2016\_09\_076\_z0.01 & 73 \\
        det\_2010\_1.15 & 81 & def\_2021\_r10\_d4.0\_z & 81 \\
        ddt\_2013\_n3 & 79 & def\_2021\_r60\_d2.6\_2z & 81 \\
        def\_2021\_r65\_d2.0\_z & 84 & gcd\_2021\_r45\_d6.0 & 79 \\
        def\_2014\_n40def & 95 & doubledet\_2012\_eldd-l & 67 \\
        def\_2021\_r129\_d4.0\_z & 75 & doubledet\_2021\_m0810\_1 & 65 \\
        ddt\_2013\_n10 & 89 & det\_2010\_1.06 & 78 \\
        det\_2010\_0.88 & 71 & ddt\_2013\_n1 & 82 \\
        def\_2021\_r45\_d6.0\_z & 82 & def\_2021\_r51\_d4.0\_z & 66 \\
        \hline
    \end{tabular}
    \caption{The SN Ia ejecta structures used in the training data set. }
    \label{tab:trainset}
\end{table}

\clearpage

\begin{table}[htb!]
    \centering
    \begin{tabular}{c|c||c|c}
        \hline
        Model Name & Number of Shells & Model Name & Number of Shells \\
        \hline
        \hline
        def\_2021\_r60\_d2.0\_z\_rot2 & 74 & ddt\_2014\_rpc40\_ddt8\_n100 & 98 \\
        def\_2021\_r48\_d5.0\_z & 78 & ddt\_2013\_n300c & 49 \\
        gcd\_2021\_r48\_d5.0 & 82 & def\_2021\_r10\_d2.0\_z & 92 \\
        doubledet\_2021\_m1002\_1 & 85 & det\_2010\_1.06\_0.075ne & 78 \\
        doubledet\_2021\_m1105\_1 & 85 & gcd\_2021\_r51\_d4.0 & 80 \\
        doubledet\_2021\_m0910r\_1 & 64 & def\_2021\_r120\_d5.0\_z & 65 \\
        def\_2021\_r10\_d5.0\_z & 85 & def\_2014\_n10def & 90 \\
        ddt\_2013\_n5 & 86 & gcd\_2021\_r57\_d3.0 & 84 \\
        def\_2014\_n3def & 90 & det\_2015\_one17e7 & 80 \\
        def\_2014\_n200def & 95 & def\_2021\_r60\_d2.6\_0.0001z & 77 \\
        \hline
    \end{tabular}
    \caption{The SN Ia ejecta structures used in the validating data set. }
    \label{tab:valiset}
\end{table}

\begin{table}[htb!]
    \centering
    \begin{tabular}{c|c||c|c}
        \hline
        Model Name & Number of Shells & Model Name & Number of Shells \\
        \hline
        \hline
        det\_2010\_0.81 & 71 & det\_2015\_one10e7 & 84 \\
        doubledet\_2012\_eldd-s & 66 & def\_2021\_r60\_d2.6\_0.001z & 81 \\
        doubledet\_2021\_m0803\_1 & 88 & doubledet\_2021\_m1003\_1 & 82 \\
        ddt\_2013\_n100\_z0.01 & 93 & doubledet\_2020\_2a\_i55 & 76 \\
        doubledet\_2020\_2a\_13 & 63 & def\_2021\_r60\_d2.6\_0.01z & 80 \\
        def\_2014\_n100hdef & 97 & def\_2021\_r60\_d2.6\_z\_co0.28 & 69 \\
        doubledet\_2021\_m1010\_1 & 86 & gcd\_2021\_r82\_d1.0 & 86 \\
        def\_2021\_r10\_d1.0\_z & 78 & def\_2021\_r10\_d2.6\_z & 82 \\
        def\_2021\_r114\_d6.0\_z & 49 & gcd\_2021\_r10\_d2.0 & 81 \\
        doubledet\_2012\_csdd-l & 60 & & \\
        \hline
    \end{tabular}
    \caption{The SN Ia ejecta structures used in the testing data set. }
    \label{tab:testset}
\end{table}

We use {\tt\string adam} \citep{Kingma2014Adam} optimization algorithm to train APNN, and we use the mean square error (MSE) as a loss function. 
The training batch size is 64, and the learning rate is 0.0001. 
The neural network is trained with 200 epochs. 
After each epoch, the neural network uses the validating data set to measure the MSE. 
The neural network is programmed with {\tt\string pytorch} framework \citep{paszke2017pytorch}, and the training time is $\sim$2 days using a NVIDIA-A100 GPU. 

\begin{figure*}
    \centering
    \includegraphics[width=0.30\textwidth]{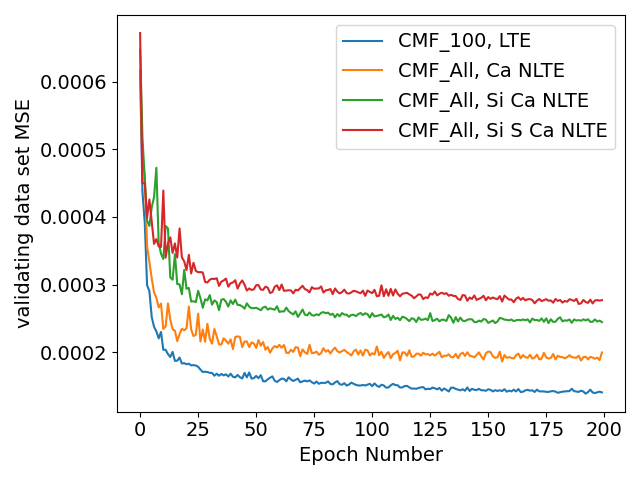}
    \includegraphics[width=0.34\textwidth]{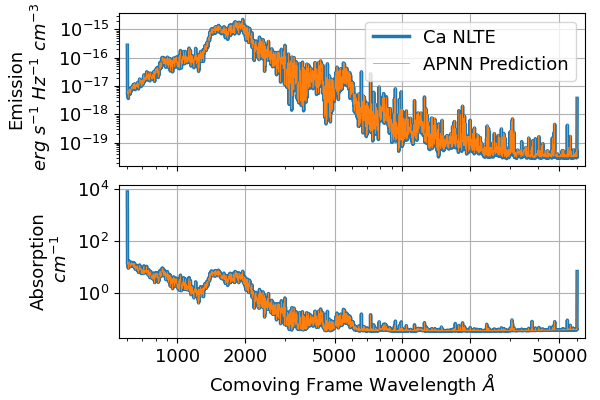}
    \includegraphics[width=0.34\textwidth]{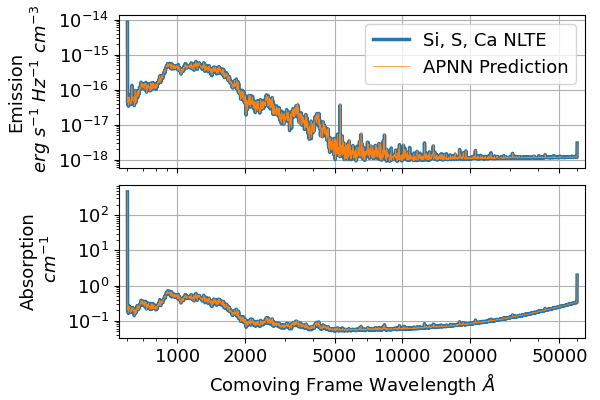}
    \caption{Left Panel: The MSE value on the validating data set of the APNNs trained on different atomic physics libraries and approximations. 
    Middle and Right Panel: A comparison of the absorption and emission coefficients calculated from APNN (orange thin line) and from traditional methods (blue thick line), note that the orange thin line and the blue thick line are overlapped. 
    Middle panel uses CMF\_All atomic library and Ca element is considered for NLTE effect. 
    Right panel uses CMF\_All atomic library and Si, S, Ca elements are considered for NLTE effect. 
    The two examples are randomly drawn from the validating data set. }
    \label{fig:aecomp}
\end{figure*}

Four APNNs are trained to account for the four atomic physics approximation recipes discussed in Section \ref{sec:nlte}. 
The left panel of Figure \ref{fig:aecomp} shows the MSE in the validating data set of the APNN trained in the data sets with different atomic physics approximations. 
With an increase in the size of the atomic data library and the inclusion of more elements for the NLTE calculation, the final MSE also increases. 
This phenomenon suggests that the increased complexity in atomic physics calculation also increases the fitting difficulty for APNN. 


In the middle and right panels of Figure \ref{fig:aecomp}, two examples of the APNN predicted absorption and emission coefficients in the validating data set are shown and compared to the truth. 
Note that the large fluctuations and the spikes in the absorption and emission coefficients are directly related to the spectral lines instead of the Monte-Carlo noise. 
We notice the APNN predictions on all the atomic physics approximation recipes are consistent with the results from the traditional method. 
In the most complex recipe (using CMF\_All atomic data library and Si, S, Ca elements considered for the NTLE effect), the performance does not decrease compared to the other three simpler recipes. 

\section{Test on 1-D Models}\label{sec:1d}

The APNN is integrated into SEDONA-GesaRaT to replace the traditional atomic physics calculation module. 
The APNN is translated into the C++ language using LibTorch \footnote{https://pytorch.org/cppdocs/index.html} module to adapt the SEDONA code. 
Thereafter, we run SEDONA-GesaRaT on the testing data set for the 1-D time-dependent RT calculation. 

\begin{figure*}
    \includegraphics[width=0.245\textwidth]{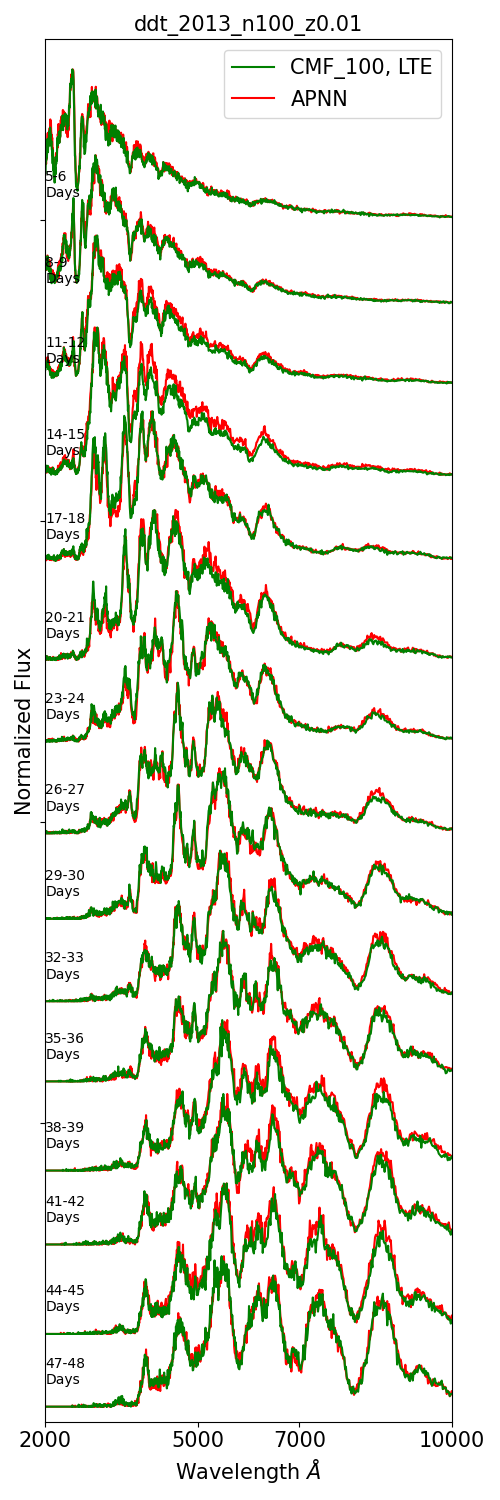}
    \includegraphics[width=0.245\textwidth]{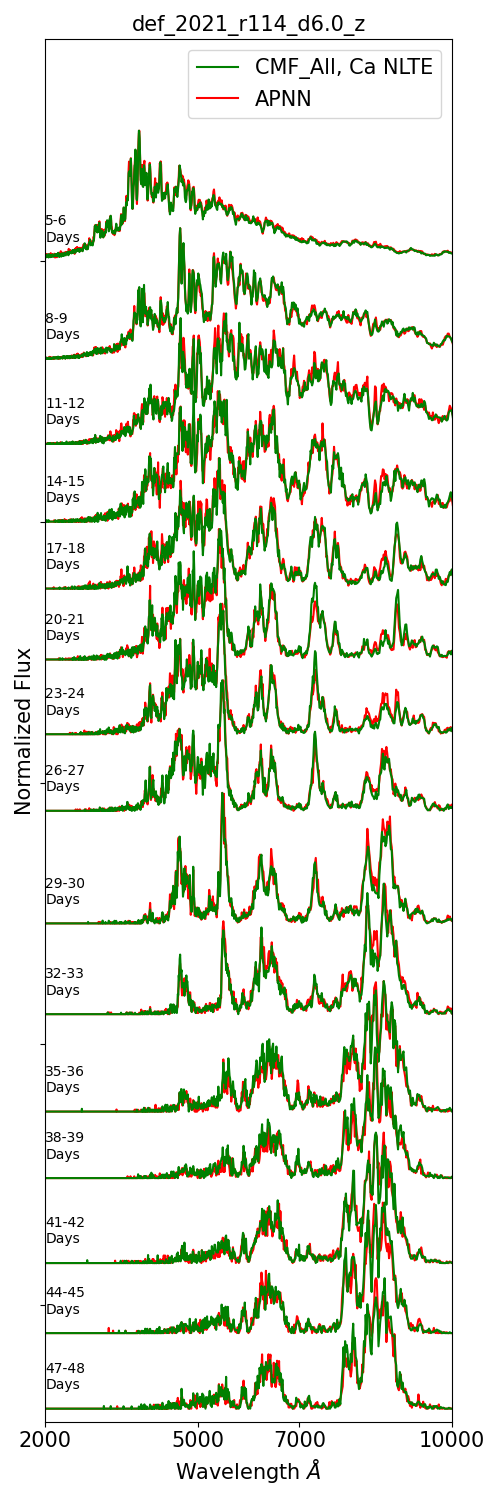}
    \includegraphics[width=0.245\textwidth]{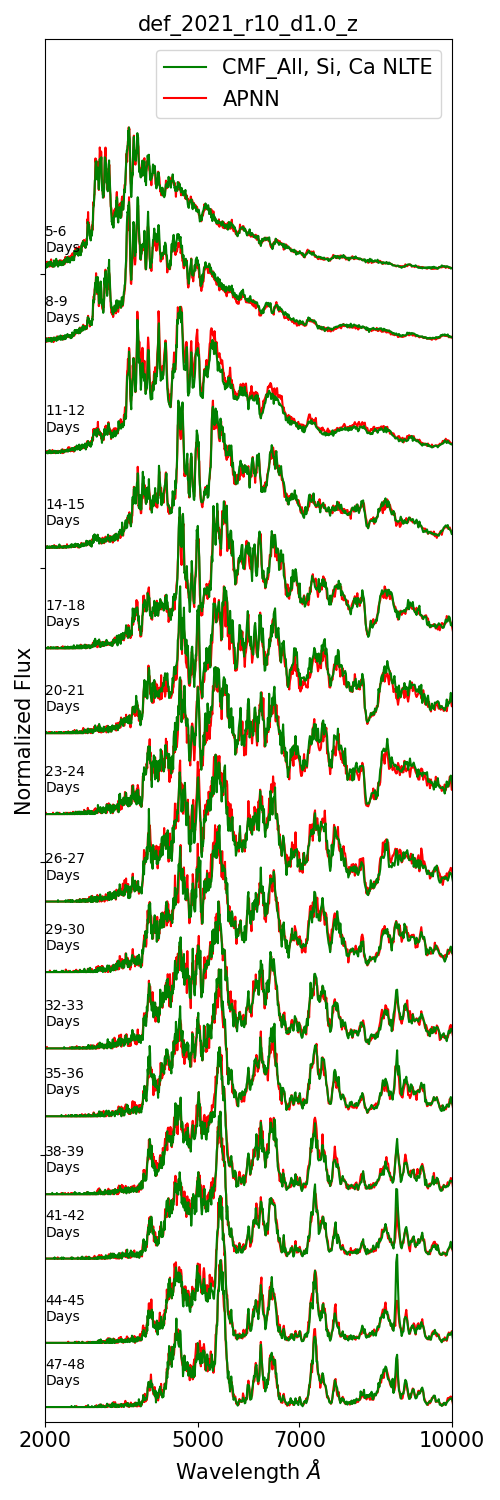}
    \includegraphics[width=0.245\textwidth]{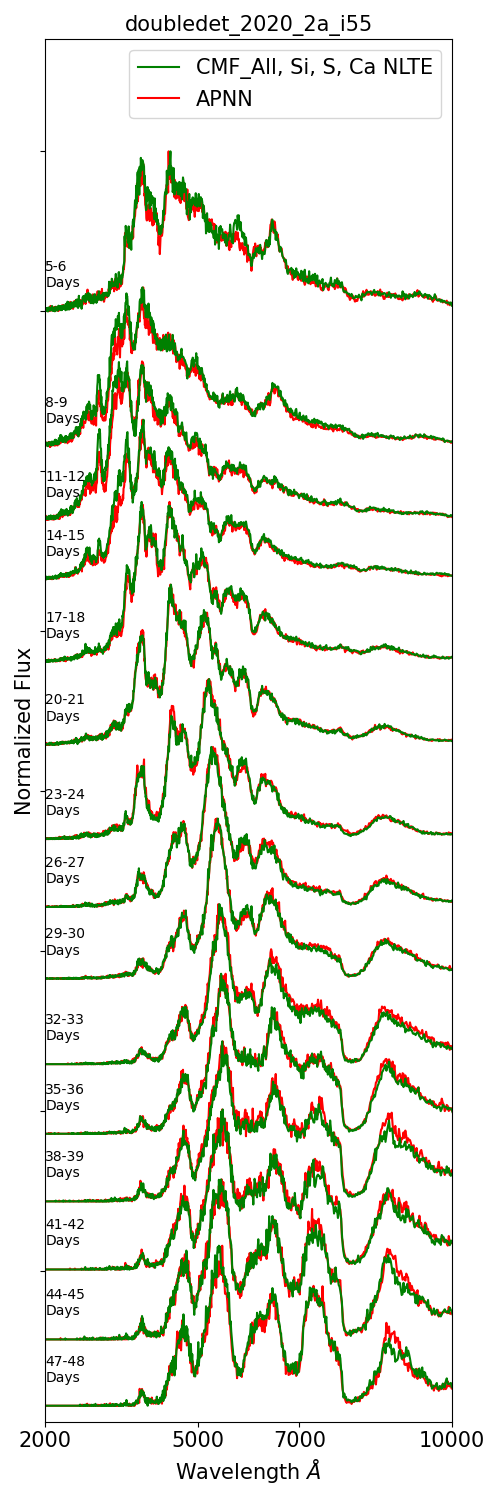}
    \caption{A comparison between the SEDONA spectral time series (green line) and SEDONA-GesaRaT spectral time series (red line). The 1-D SN Ia explosion model names are shown in the title of each panel. The time after the explosion is labeled on the left of each spectrum. }
    \label{fig:1dspec}
\end{figure*}

Figure \ref{fig:1dspec} shows several examples of comparing the SEDONA-GesaRaT spectra and the original SEDONA spectra using the SNe Ia models in the testing data set. 
The SEDONA-GesaRaT spectra have successfully reproduced the spectral time sequence of the original SEDONA from early to late phase, and the spectral features are captured with high accuracy. 

Table \ref{tab:apnntime} shows the typical computation time measured on one core of an Intel Xeon 6248R CPU. 
Note that the computation time of the APNN only depends on the neural network structure, and the four APNNs trained on different atomic recipes share the same computation time. 

\begin{table}[htb!]
    \centering
    \begin{tabular}{c|c}
        \hline
        Atomic recipes & run time (core-seconds) \\
        \hline
        \hline
        CMF\_100 LTE            & 1.19  \\
        CMF\_All Ca NLTE        & 57.92 \\
        CMF\_All Si, Ca NLTE    & 65.86 \\
        CMF\_All Si, S, Ca NLTE & 72.09 \\
        APNN                    & \textbf{0.17}  \\
        \hline
    \end{tabular}
    \caption{The computation time of solving the level population, calculating absorption and emission coefficients in one zone using four different atomic libraries and approximations, or using APNN. }
    \label{tab:apnntime}
\end{table}

\end{document}